\documentclass[10pt,conference]{IEEEtran} 
\IEEEoverridecommandlockouts
\usepackage{balance}
\usepackage[table]{xcolor}
\usepackage{xcolor}
\usepackage{float}
\usepackage{listings}
\usepackage{soul}
\usepackage[numbers]{natbib}
\definecolor{codegreen}{rgb}{0,0.6,0}
\definecolor{codegray}{rgb}{0.5,0.5,0.5}
\definecolor{codepurple}{rgb}{0.58,0,0.82}
\definecolor{backcolour}{rgb}{0.95,0.95,0.92}
\lstdefinestyle{mystyle}{
    backgroundcolor=\color{backcolour},   
    commentstyle=\color{codegreen},
    keywordstyle=\color{magenta},
    numberstyle=\tiny\color{codegray},
    stringstyle=\color{codepurple},
    basicstyle=\ttfamily\footnotesize,
    breakatwhitespace=false,         
    breaklines=true,                 
    captionpos=b,                    
    keepspaces=true,                 
    numbers=left,                    
    numbersep=5pt,                  
    showspaces=false,                
    showstringspaces=false,
    showtabs=false,                  
    tabsize=2
}

\usepackage[skip=0.333\baselineskip]{caption}
\usepackage{tabularx}

\usepackage{amsmath}
\usepackage{array}
\usepackage{xcolor}
\usepackage{tcolorbox}
\usepackage{listings}
\usepackage{algorithm}
\usepackage{algpseudocode}
\usepackage{amsmath,amssymb,amsfonts}
\usepackage{titlesec}
\usepackage{xspace}
\usepackage{enumitem}
\usepackage{pdfpages}
\usepackage{pgfplots}
\usepackage{soul}
\usepackage{url}
\usepackage{hyperref}
\usepackage[normalem]{ulem}

\usepackage{multirow}

\newcommand{\peter}[1]{\textcolor{red}{{\it [peter says: #1]}}}
\newcommand{\feng}[1]{\textcolor{olive}{{\it [feng says: #1]}}}
\usepackage{booktabs}
\usepackage{cancel}
\usepackage{tikz}
\usetikzlibrary{calc}
\usepackage[svgpath=../Figure/]{svg}

\newlength\Linewidth

\newcommand{\raw}{\textit{RawGPT}\xspace}
\newcommand{\codeT}{\textit{CodeT}\xspace}
\newcommand{\reflexion}{\textit{Reflexion}\xspace}
\newcommand{\water}{\textit{Waterfall}\xspace}
\newcommand{\tdd}{\textit{Test-Driven-Development (TDD)}\xspace}
\newcommand{\tdds}{\textit{TDD}\xspace}
\newcommand{\scrum}{\textit{Scrum}\xspace}
\newcommand{\scrumTest}{\textit{Scrum+Test}\xspace}

\newcommand{\phead}[1]{\vspace{1mm} \noindent {\bf #1}}

\newcommand{\tool}{\textit{FlowGen}\xspace}
\newcommand{\delete}[1]{\textcolor{red}{\sout{}}}
\newcommand{\add}[1]{{#1}}

\newcommand{\toolWater}{$\tool_{\water}$\xspace}
\newcommand{\toolTdd}{$\tool_{\tdds}$\xspace}
\newcommand{\toolScrum}{$\tool_{\scrum}$\xspace}
\newcommand{\toolScrumTest}{$\tool_{\scrumTest}$\xspace}

\newcommand{\framework}{\textbf{Flow-LLM-CodeGen}\xspace}

\newcommand{\rqboxc}[1]{\begin{tcolorbox}[left=1pt,right=1pt,top=1pt,bottom=1pt,colback=gray!35,colframe=gray!35,before skip=4pt,after skip=1pt]#1\end{tcolorbox}}
\titlespacing*{\subsubsection}{0pt}{2pt}{0.5pt}

\newcommand\jsonkey{\color{purple}}
\newcommand\jsonvalue{\color{cyan}}
\newcommand\jsonnumber{\color{orange}}

\makeatletter
\newif\ifisvalue@json

\lstdefinelanguage{json}{
    tabsize             = 4,
    showstringspaces    = false,
    keywords            = {false,true},
    alsoletter          = 0123456789.,
    morestring          = [s]{"}{"},
    stringstyle         = \jsonkey\ifisvalue@json\jsonvalue\fi,
    MoreSelectCharTable = \lst@DefSaveDef{`:}\colon@json{\enterMode@json},
    MoreSelectCharTable = \lst@DefSaveDef{`,}\comma@json{\exitMode@json{\comma@json}},
    MoreSelectCharTable = \lst@DefSaveDef{`\{}\bracket@json{\exitMode@json{\bracket@json}},
    basicstyle          = \ttfamily \footnotesize
}

\newcommand\enterMode@json{%
    \colon@json%
    \ifnum\lst@mode=\lst@Pmode%
        \global\isvalue@jsontrue%
    \fi
}

\newcommand\exitMode@json[1]{#1\global\isvalue@jsonfalse}

\lst@AddToHook{Output}{%
    \ifisvalue@json%
        \ifnum\lst@mode=\lst@Pmode%
            \def\lst@thestyle{\jsonnumber}%
        \fi
    \fi
    \lsthk@DetectKeywords%
}

\makeatother

 \pgfplotsset{compat=1.18}
\begin{document}



\title{SOEN-101: Code Generation by Emulating Software Process Models Using Large Language Model Agents}

\author{
    \IEEEauthorblockN{Feng Lin\textsuperscript{1}, Dong Jae Kim\textsuperscript{2}, Tse-Hsun (Peter) Chen\textsuperscript{1}}%
    \IEEEauthorblockA{\textit{\textsuperscript{1}Software PErformance, Analysis, and Reliability (SPEAR) lab, Concordia University, Montreal, Canada}}
    \IEEEauthorblockA{\textit{\textsuperscript{2}DePaul University, Chicago, USA}}

    \IEEEauthorblockA{feng.lin@mail.concordia.ca, k\_dongja@encs.concordia.ca,  peterc@encs.concordia.ca}
}
\maketitle

\begin{abstract} 
Software process models are essential to facilitate collaboration and communication among software teams to solve complex development tasks. Inspired by these software engineering practices, we present \tool\delete{ in this paper }-- a code generation framework that emulates software process models based on multiple Large Language Model (LLM) agents. We emulate three process models, \toolWater, \toolTdd, and \toolScrum, by assigning LLM agents to embody roles (i.e., requirement engineer, architect, developer, tester, and scrum master) that correspond to everyday development activities and organize their communication patterns. The agents work collaboratively using chain-of-thought and prompt composition with continuous self-refinement to improve the code quality. We use GPT3.5 as our underlying LLM and \add{several} baseline\add{s} (\raw, \add{\codeT, \reflexion}) \add{to evaluate code generation on four benchmarks:} \delete{and conduct the evaluation on four code generation benchmarks }HumanEval, HumanEval-ET, MBPP, and MBPP-ET. 
Our findings show that \toolScrum excels compared to other process models, achieving a Pass@1 of 75.2, 65.5, 82.5, and 56.7 in HumanEval, HumanEval-ET, MBPP, and MBPP-ET, respectively (an average of 15\% improvement over \raw). 
\delete{Our analysis also reveals that different development activities have different impacts on code smell and exception-handling ability in the generated code.}
\add{Compared with other state-of-the-art techniques, \toolScrum achieves a higher Pass@1 in MBPP compared to \codeT, with both outperforming \reflexion. Notably, integrating \codeT into \toolScrum resulted in statistically significant improvements, achieving the highest Pass@1 scores. Our analysis also reveals that the development activities impacted code smell and exception handling differently, with design and code review adding more exception handling and reducing code smells. Finally, \tool models maintain stable Pass@1 scores across GPT3.5 versions and temperature values, highlighting the effectiveness of software process models in enhancing the quality and stability of LLM-generated code.} 
\delete{In particular, design and code review help add more exception-handling code, while design, testing, and code review help reduce code smells. Finally, we find that the temperature values do not significantly affect Pass@1 in all models, but Pass@1 of \raw can be substantially impacted if the GPT3.5 model version changes (e.g., Pass@1 ranges from 5 to over 60 in HumanEval). In contrast, all three models achieve a stable Pass@1 across model versions, showing the stability of \tool. In conclusion, the findings underscore the importance of adopting software process models to enhance the quality and stability of LLM-generated code.}

\end{abstract}


\section{Introduction}
The recent surge of Large Language Models (LLMs) has sparked a transformative phase in programming and software engineering. Pretrained on vast repositories of code-related datasets, these LLMs have acquired a comprehensive understanding of code, enabling them to excel in diverse code-related tasks. 
With tools like ChatGPT~\cite{chatGPT} or LLaMA~\cite{LLAMA}, researchers have demonstrated the potential of LLMs in generating commit messages~\cite{zhang2024automatic}, resolving merge conflicts~\cite{shen2023git}, generating tests~\cite{xie2023chatunitest,yuan2023no,schafer2023empirical}, method renaming~\cite{alomar2024refactor}, and even facilitating log analytics~\cite{ma2024knowlog,ma2024llmparser}.

Among all development activities, code generation has received much attention due to its potential to reduce development costs. As LLMs are becoming increasingly integral to software development, various techniques have emerged in LLM-based code generation. For example, prompting techniques like few-shot learning~\cite{kumar2021reordering,yuan2023evaluating} have been shown to improve code generation results. In particular, few-shot learning coupled with few-shot sampling~\cite{ma2024llmparser,kang2023large} or information retrieval augmented technique~\cite{nashid2023retrieval,chen2023benchmarking} have been shown to improve code generation. 
Moreover, one can integrate personalization in the prompt, instructing LLMs to be domain experts in a specific field, which can further improve LLM responses~\cite{white2023chatgpt,shao2023character}. Such personalization techniques highlight the potential of using multiple LLMs working together to assist in complex software development activities.


Given the complexity of software development, LLM agents stand out among various LLM techniques. Agents are LLM instances that can be customized to carry out specific tasks that replicate human workflow \cite{hong2023metagpt,dong2023self}. Recently, multi-agent systems have achieved significant progress in solving complex problems in software development by emulating development roles~\cite{hong2023metagpt,dong2023self,qian2023communicative}. 
MetaGPT, introduced by \citet{hong2023metagpt}, integrated development workflow using standard operating procedures by assigning specific roles (e.g., a designer or a developer) to LLM agents. \citet{dong2023self} developed self-collaboration, which assigns LLM agents to work as distinct ``\textit{experts}'' for sub-tasks in software development. \citet{qian2023communicative} proposed an end-to-end framework for software development through self-communication among the agents. 

Despite the promising applications of LLMs in automating software engineering tasks, it is pivotal to recognize that software development is a collaborative and multi-faceted endeavor. In \delete{the real world}\add{practice}, developers and stakeholders work together, following certain software process models like \water, \tdd, and \scrum. The process models help facilitate communication and collaboration to ensure the delivery of high-quality products. 
Even though there is a common community agreement on the pros and cons of each process model~\cite{processTradeOffs}, the impact of adopting these process models for LLM code generation tasks remains unknown. In particular, will emulating different process models impact the generated code quality in different aspects, such as reliability, code smell, and functional correctness?

While some research has explored integrating multi-agents within LLM frameworks~\cite{qian2023communicative, xu2023exploring, yang2023auto}, their research focus diverges from the influence of the software process model on code generations for several reasons: 1)~\citet{xu2023exploring} do not adhere to specific process models, and 2) both \citet{dong2023self} and \citet{qian2023communicative} focus solely on \textit{Waterfall}\add{\textit{-like} models}, neglecting \textit{TDD} and \scrum, which may have different impact on code generations. Importantly, none of the aforementioned studies conduct a fine-grain analysis of how different development activities affect code quality metrics, such as code smell and reliability, other than the Pass@1 score. Our study takes further steps to analyze the impacts of different agents within the process models on code generation and their influence on other code quality attributes.

This paper presents a novel multi-agent \delete{Large Language Model (LLM) }\add{LLM-}based code generation framework named \tool. \tool integrates diverse prompt engineering techniques, including chain-of-thought~\cite{wei2022chain, li2023structured}, prompt composition~\cite{10.1145/3560815, yuan2023no}, and self-refinement~\cite{madaan2024self}, with a focus on emulating the flow of development activities in various software process models. Specifically, we implemented three popular process models into \tool: \toolWater, \toolTdd, and \toolScrum. Each process model emulates a real-world development team involving several LLM agents whose roles (i.e., requirement engineer, architect, developer, tester, and scrum master) correspond to common software development activities. The agents \delete{are assigned roles with different domain-specific expertise, working}\add{work} collaboratively to produce software artifacts and help other agents review and improve the artifacts in every activity.\delete{ of a process model.} 

We evaluate \tool on four popular code generation benchmarks: \textit{HumanEval}~\cite{chen2021evaluating}, \textit{HumanEval-ET}~\cite{dong2023codescore}, \textit{MBPP}~\cite{austin2021program}, and \textit{MBPP-ET}~\cite{liu2023your}. We apply zero-shot learning to avoid biases in selecting few-shot samples~\cite{xu2022alleviating}. To compare, we also apply zero-shot learning on GPT-3.5 as our baseline (\raw). We repeat our experiments five times to account for variability in LLM's responses and report the average value and standard deviation. To study code quality, in addition to Pass@1, we run static code checkers to detect the prevalence of code smells in the generated code. 
Our evaluation shows 
that \toolScrum's generated code achieves the highest accuracy (Pass@1 is 75.2 for HumanEval, 65.5 for HumanEval-ET, 82.5 for MBPP, and 56.7 for MBPP-ET), surpassing \raw's Pass@1 by 5.2\% to 31.5\%. 
While \tool, in general, is more stable than \raw, \toolScrum exhibits the most stable results with an average standard deviation of only 1.3\% across all benchmarks. 

\add{Additionally, we compare \toolScrum with other state-of-the-art techniques: \codeT~\cite{chen2022codet} and \reflexion~\cite{shinn2024reflexion}. Both \toolScrum and \codeT outperform \reflexion significantly in terms of Pass@1 across all benchmarks, with \toolScrum achieving a higher Pass@1 than \codeT in MBPP. Furthermore, the integration of \codeT into \toolScrum demonstrates the highest Pass@1 scores, highlighting the potential of integrating other prompting techniques with \tool for improved code generation.}

We further study the impact of each development activity on code quality. We find that removing the testing activity in the process model results in a significant decrease in Pass@1 accuracy (17.0\% to 56.1\%). Eliminating the testing activity also leads to a substantial increase in error and warning code smell densities. We also find that the design and code review activities reduce refactor and warning code smells, and improve reliability by adding more exception handling code. 
Nevertheless, \tool consistently outperforms \raw by reducing code smells and enhancing exception handling. Finally, we find that the GPT model version plays a significant role in the quality of generated code, and \tool helps ensure stability across different versions of LLMs and temperature values. 


We summarize the main contributions of this paper as follows:

\begin{enumerate}
    \item \textbf{Originality}: We introduce a multi-agent framework called \tool, incorporating software process models from real-world development practice. We integrate agents acting as requirement engineers, architects, developers, testers, and scrum masters, and study how their interaction improves code generation and code quality. 
    
    \item \textbf{Technique}: We integrate prompt engineering techniques like chain-of-thought, prompt composition, and self-refinement to facilitate interactions among the agents. We implement three recognized process models: \toolWater, \toolTdd, and \toolScrum, but the technique can be easily extended to emulate other process models or development practices (e.g., DevOps). 

    \item \textbf{Evaluation}: We conduct a fine-grained evaluation on the quality of the generated code using four popular code generation benchmarks: \textit{Humaneval}~\cite{chen2021evaluating}, \textit{Humaneval-ET}~\cite{dong2023codescore}, \textit{MBPP}~\cite{austin2021program}, \textit{MBPP-ET}~\cite{liu2023your}, comparing agent interactions and their effect on both accuracy (Pass@1) and other code quality metrics (e.g., smells). We manually checked the generated code and discussed the reasons for test failures. Finally, we examined how model versions and temperature settings affect code generation stability. 

    \item \textbf{Data Availability}: To encourage future research in this area and facilitate replication, We made our data and code publicly available online~\cite{dataset}. 


\end{enumerate}

\noindent \textbf{Paper Organization.} Section~\ref{background} discusses background and related work. Section~\ref{method} provides the details of \tool. Section~\ref{result} evaluates our \tool. \add{Section~\ref{sec:discussion} provides a discussion on future work.} Section~\ref{threats} discusses threats to validity. Section~\ref{conclusion} concludes the paper.

\section{Background \& Related Works}\label{background}
In this section, we discuss the background of software process models and LLM agents. We also discuss related work on LLM-based code-generation. 

\subsection{Background}

\noindent\textbf{Software Development Process.}
Software development processes encompass methodologies and practices that development teams use to plan, design, implement, test, and maintain software. 
The primary goal of a software process is to assist the development teams in producing high-quality software. 
Generally, different software process models involve the same set of development activities, such as requirement, design, implementation, and testing, but differ in how the activities are organized. 
Because of the variation, each software process model has its strengths and weaknesses based on the project type, teams, and experience~\cite{processTradeOffs}. 

In particular, three well-known and widely adopted software process models were created over the years: \water, \tdd, and \scrum.
\water \cite{bassil2012simulation} is often used in safety-critical systems where development teams must adhere to a linear path, and each software development activity builds upon the previous one. 
\tdds and \scrum are both variants of the agile development model. Compared to \water, agile process models focus more on iterative and incremental development and adapting to change. 
\tdds \cite{maximilien2003assessing} emphasize writing tests before writing the actual code to improve software design and quality. 
\scrum highlights the importance of collaboration and communication in software development. \scrum prescribes for teams to break work into time-boxed iterations called sprints. During these sprints, teams focus on achieving specific goals (e.g., user stories), ensuring a continuous discussion among teams to handle any unexpected risks throughout the development process.


\noindent\textbf{LLM Agents.} 
LLM agents are artificial intelligence systems that utilize LLM as their core computational engines to understand questions and generate human-like responses. 
LLM agents can refine their responses based on feedback, learn from new information, and even interact with other AI agents to collaboratively solve complex tasks \cite{hong2023metagpt,qian2023communicative,xu2023exploring,park2023generative}. Through prompting, agents can be assigned different roles (e.g., a developer or a tester) and provide more domain-specific responses that can help improve the answer \cite{hong2023metagpt, white2023chatgpt, shao2023character}. 

One vital advantage of agents is that they can be implemented to interact with external tools. When an agent is reasoning the steps to answer a question, it can match the question/response with corresponding external tools or APIs to construct or refine the response. For instance, an LLM agent that represents a data analysis engineer can apply logical reasoning to generate corresponding SQL query statements, invoke the database API to get the necessary data, and then answer questions based on the returned result. 
When multiple LLM agents are involved, they can collaborate and communicate with each other. Such communication is essential for coordinating tasks, sharing insights, and making collective decisions. Hence, defining how the agents communicate can help optimize the system's overall performance~\cite{xi2023rise}, allowing agents to undertake complex projects by dividing tasks according to their domain-specific skills or knowledge.

The software development process plays a crucial role in \add{software development}\delete{the creation of high-quality software}, fundamentally involving communication among various development roles. 
Given the demonstrated capability of Large Language Model (LLM) agents to mimic domain experts in specific fields \cite{white2023chatgpt, hong2023metagpt, qian2023communicative}, this study leverages LLM agents to represent diverse development roles and conduct their associated duties. 
Our research establishes a collaborative team of LLM agents designed to emulate these process models and roles, aiming to enhance code generation.




\subsection{Related Works}

Code generation is a thriving field of research because of its potential to reduce development costs. \delete{Apart from training more powerful LLM to improve code generation, }\add{In particular, }\textit{prompt-based} and \textit{agent-based} code generation techniques are two of the most prevalent directions.

\noindent\textbf{Prompt-based Code Generation.} 
Prompt-based code generation employs a range of techniques to refine prompts, ultimately leading to the generation of expected code. For example, \citet{li2023structured} propose using structured prompts containing code information (e.g., branch and loop structures) to improve the generated code. \citet{nashid2023retrieval} retrieval code demos similar to the given task and include them in the prompt to improve code generation. 
\citet{ruiz2024novel} use translation techniques for program repair, where buggy programs are first translated into natural language or other programming languages. The translated code is used as a prompt to generate new/fixed code with the same feature. 
\citet{schafer2023empirical} iteratively refine prompts based on feedback received from interpreters or test execution results. \citet{kang2023large} provide specific instructions, test method signature, and bug report as part of the prompt for generating test code to reproduce bugs. \citet{xie2023chatunitest} parse the code to identify the focal method and related code context, which are given in the prompt for test code generation. 
\citet{yuan2023no} apply a prompt composition technique by first asking an LLM to provide a high-level description of a method, and then the description is used as part of the prompt to enhance test code generation. \add{\citet{chen2022codet} introduced \codeT, a framework that employs self-generated tests to evaluate the quality of generated code. \citet{shinn2024reflexion} presented \reflexion, which utilizes an evaluator LLM to provide feedback for enhancing future decision-making processes. }





\noindent\textbf{Agent-based Code Generation.} 
Agent-based code generation emphasizes on the importance of role definition and communication among multiple LLM agents. Some approaches incorporate external tools as agents. For example, \citet{huang2023agentcoder} introduce the test executor agent, employing a Python interpreter to provide test logs for LLMs. \citet{zhong2024ldb} introduces a debugger agent that utilizes a static analysis tool to build control flow graph information, guiding LLMs in locating bugs. 
Meanwhile, other studies~\cite{hong2023metagpt,qian2023communicative,dong2023self} task LLMs as agents by emulating diverse human roles, including analysts, engineers, testers, project managers, chief technology officers (CTOs), etc. 
\add{Nevertheless, these studies miss key roles in the development activities (e.g., only has analysts, coders, and testers~\cite{dong2023self}) or focus more on the business side of the roles (e.g., employ CTO and CEO)~\cite{qian2023communicative}. In our work, we try to follow the Waterfall model that is proposed in the software engineering literature and create agents that correspond to every development activity.} 
These approaches follow the \water model to communicate among these roles, with variation in the prompts and roles, ultimately improving code generation.

In comparison, our research leverages LLM agents to emulate multiple software development process models, while prior research focuses only on the \water model~\cite{hong2023metagpt,qian2023communicative,dong2023self}. We implement several prompting techniques, but more importantly, we emphasize on how various process models and the associated development activities affect the generated code. Different from prior works which only study functional correctness, we study several additional dimensions of code quality, including code design, code smell, convention issues, and reliability. We also explore why the generated code fails the tests and the sensitivity of the results across LLM model versions and temperature values. \delete{
While there are many code generation results using the same benchmarks, our evaluation results cannot be directly compared with some other agent-based code generation research~\cite{hong2023metagpt,huang2023codecot, huang2023agentcoder,chen2022codet,shinn2024reflexion} due to missing information on the model version, temperature value, steps to post-process the generated code, specific prompts, or how the few-shot samples are selected. Nevertheless, \tool still shows improved Pass@1 compared to existing works that provide clear information. for instance, \citet{dong2023self} (uses the same GPT3.5 version as our study) applied few-shot learning with LLM agents and received a Pass@1 of 68.2\% in MBPP. In contrast, while \toolScrum uses zero-shot learning, we achieved a Pass@1 of 82.5\% in MBPP. We observe the same level of improvements in all the benchmarks. Similarly, \citet{li2023structured}, which used the same model version and temperature value as we do, applied structured chain-of-thought to enhance reasoning ability and achieved a Pass@1 of 60.6\% in HumanEval. 
} \delete{\toolScrum attained 78.5\% under the same model version and temperature setting. In short, due to variations of the LLMs and missing information in prior works, we focus on comparing the \tool's improvement against \raw (directly using GPT3.5) in this study.} 
\section{Methodology}\label{method}
\label{sec:methodlogy}

\begin{figure*}
    \centering
\includegraphics[width=1\textwidth]{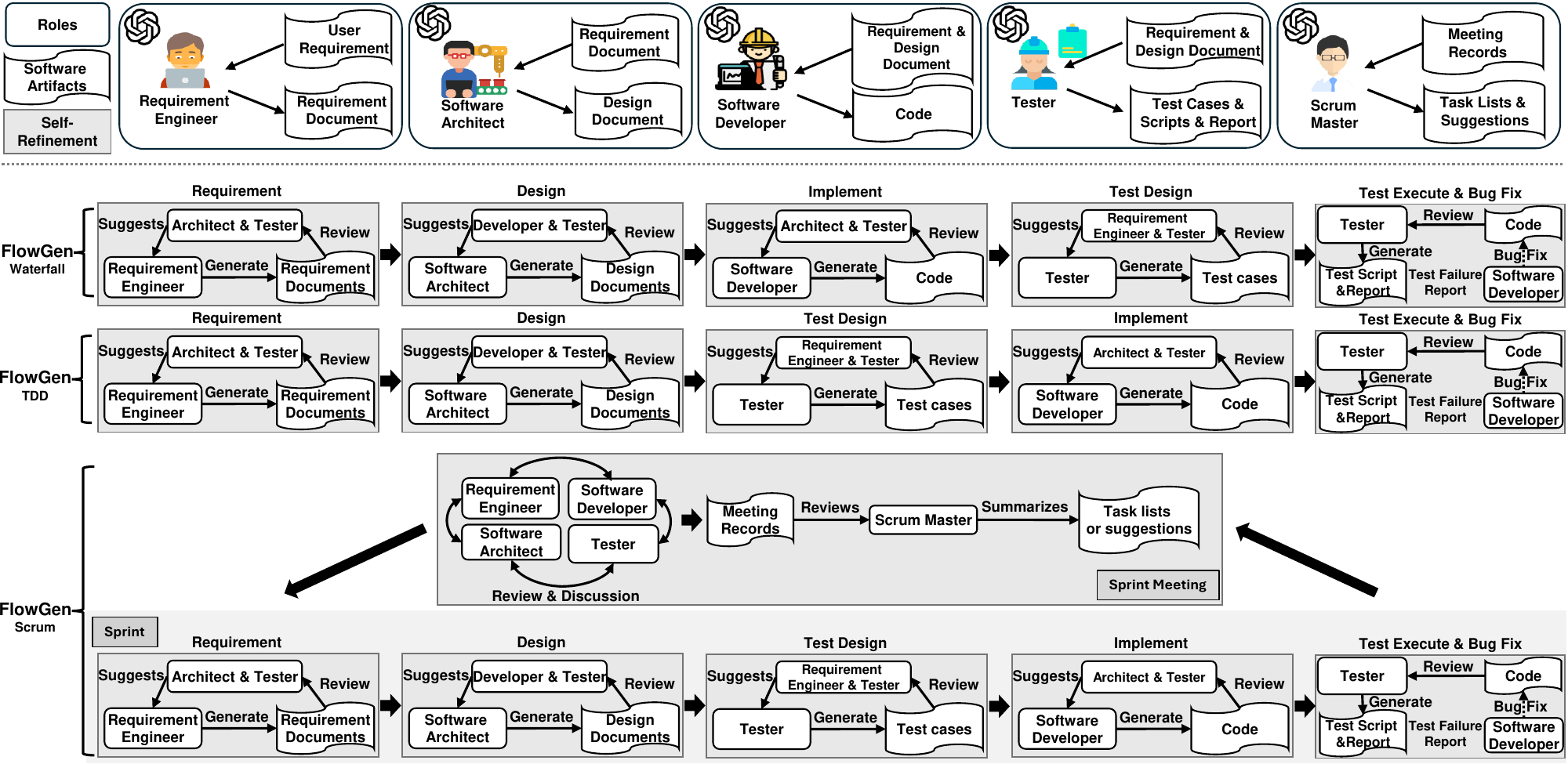}
    \caption{An overview of \toolWater, \toolTdd, and \toolScrum.}
    \label{figures: overview}
\end{figure*}

We propose \tool, an agent-based code generation technique based on emulating different software processes. Figure~\ref{figures: overview} shows the overview of \tool: (1) define the roles and their responsibilities; (2) use LLM agents to represent these roles; and (3) complete the interactions among these agents according to the software process models. 
In each development activity, we implement chain-of-thought and self-refinement to improve the quality of the generated artifacts. 
In particular, we study and compare three software process models: {\em Waterfall}, {\em TDD}, and {\em Scrum}. Nevertheless, \tool can be easily adapted to different process models. We use zero-shot learning in all our experiments to avoid biases in selecting data samples. Below, we discuss \tool in detail.

\subsection{Using LLM Agents to Represent Different Development Roles in Software Process Models} 
\label{sec:agents}

In \tool, we create LLM agents who are responsible for the main development activities: requirement, design, implementation, and testing. Hence, to emulate a software process model, we reorganize the communication and interaction among different agents. The benefit of such a design is that it maximizes the extensibility and \delete{reuseable}\add{reusability} of the agents, and \tool can be easily adapted to different process models. 
We implement four agents whose role corresponds to the common development activities: \textit{Requirement Engineer}, \textit{Architect}, \textit{Developer}, and \textit{Tester}. For \scrum, we introduce an additional role -- \textit{Scrum Master}. {\delete{We design these roles so that they use the same prompt template with role-specific details as shown below:} }

\add{We designed these roles to use the same prompt template across different process models (with different terms such as user stories v.s. requirement) to investigate the effectiveness of process models on code generation. The exact words that we used for the prompts can be found online~\cite{dataset}. The role-specific details of our prompt are:}

\begin{lstlisting}[language=json]
{
  "Role": "You are a [role] responsible for [task]",
  "Instruction": "According to the Context please [role-specific instruction]",
  "Context": "[context]"
}
\end{lstlisting}

In this prompt template \add{(inspired by MetaGPT~\cite{hong2023metagpt} and Self-Collaboration~\cite{dong2023self})}, {\tt role} refers to one of the roles (e.g., Requirement Engineer) that corresponds to the development activity, and {\tt task} describes the duties for the {\tt role}  (e.g., analyze and create requirement documents). {\tt Instruction} leverages chain-of-thought reasoning~\cite{wei2022chain, li2023structured} and refers to role-specific instruction listed in steps, such as 1) analyzing the requirement and 2) writing the requirement documentation. Finally, {\tt Context} contains the programming question, the agent conversation history, or the agent-generated artifacts. {\tt Context} includes all necessary information that helps the agents to make a next-step decision based on the current conversation and generated results.

Table~\ref{table:prompt} shows the {\tt tasks}, {\tt instructions}, and {\tt contexts} for every development role. In general, every role takes the output from the prior development activities as input (i.e., {\tt context}). For example, an architect writes a design document based on the requirement document generated by a requirement engineer. We design developers and testers to have multiple tasks. Developers are responsible for writing code and fixing/improving the code based on suggestions. We design testers using a prompt composition technique, which is shown to improve the LLM-generated result~\cite{10.1145/3560815, yuan2023no}. First, testers design a test. Then, testers write and execute the tests based on the design. \add{On average, \tool generates four tests for each problem before the review meeting and six after the meeting. It is important to note that oracles are kept aside and never used in the code-generation process.} Finally, testers generate a test failure report.

Developers receive the test failure report to fix the code. In addition to the tasks described in Table~\ref{table:prompt}, all roles have one common task, which is to provide feedback to other roles for further improvement (e.g., for code review). 

\begin{table*}
    \centering
    \caption{Tasks, instructions, and corresponding contexts that are used for constructing the prompts for the development roles.}
    \label{table:prompt}
    \scalebox{0.85}{
    \begin{tabular}{p{0.14\textwidth}|p{0.25\textwidth}p{0.46\textwidth}p{0.22\textwidth}}

    \toprule
    Role & Task & Instruction & Context\\
    \midrule
    Requirement Engineer  &  Analyze and generate requirement documentation from the context. &  1) Analyze the requirement; and 2) Write a requirement document. & Programming problem description. \\
    \midrule
    Architect  & Design the overall structure and high-level components of the software. & 1) Read the context documents; and 2) Write the design document. The design should be high-level and focus on guiding the developer in writing code. & Requirement document. \\
    \midrule  
    Developer & Write code in Python that meets the requirements. & 1) Read the context documents; and 2) Write the code. Ensure that the code you write is efficient, readable, and follows best practices. & Requirement and design documents. \\
     & Fix the code so that it meets the requirements. & 1) Read the test failure reports and code suggestions from the context; and 2) Rewrite the code. & Original code, test failure report, and suggestions for improvement. \\
    \midrule
    Tester & Design tests to ensure the software satisfies feature needs and quality. & 1) Read context documents; and 2) Design test cases. & Requirement and design documents. \\
    & Write a Python test script using the {\sf unittest} framework. & 1) Read the context documents; 2) Write a Python test script; and 3) Follow the input and output given by the requirement. & Test case design and requirement documents.\\
    & Write a test failure report. & 1) Read the test execution result; and 2) Analyze and generate a test failure report. & Test execution result. \\

    \midrule
    Scrum Master & Summarize and break down the discussion into a task list for the scrum team. & 1) Read and understand the context; and 2) Define the tasks for development roles. & Meeting discussion.\\
    \bottomrule
    \end{tabular}}
    \label{tab:roles_prompt_template}
    \vspace{-0.3mm}
\end{table*}

\subsection{Communications Among Agents} \label{lab: interactions}


One of the most important aspects of LLM agents is how the agents communicate. A recent survey paper~\cite{xi2023rise} shows that one common communication pattern is sequential, i.e., ordered,  where one agent communicates to the next in a fixed order. 
Another pattern is disordered, where multiple agents participate in the conversation. Each agent gets the context separately and outputs the response in a shared buffer. Then, the responses can be summarized and used in the next decision-making process. 
Based on the software process models and the two aforementioned communication patterns, we implement three interaction models for the agents: \toolWater, \toolTdd, and \toolScrum (Figure~\ref{figures: overview}). 
\add{The details of our multi-agent communication history are available online~\cite{dataset}.}

\subsubsection{\toolWater}
\add{\toolWater follows the \water model and implements}
\delete{We follow the intuition behind the \water model and implement \toolWater as} an ordered communication among the agents. Given a programming problem, the problem goes through the requirement analysis, design, implementation, and testing. One thing to note is that the test result from our generated tests is redirected to the developer agent in our implementation of \toolWater so developers can fix and improve the code. 

\subsubsection {\toolTdd}
In the design of \toolTdd, we follow the ordered communication pattern and organize the development activities so that testing happens after design and before implementation. Once the tests are written, the developer agent considers the test design when implementing the code. When the implementation is finished, we execute the tests. If a test fails, the developer agent is asked to examine the code and resolve the issue.  

\subsubsection {\toolScrum} Compared to \water and \tdds, \scrum involves one additional role, the \textit{Scrum Master}. There are also additional {\it Sprint meetings} among the agents. Note that, different from \toolWater, we use the agile terminologies in the prompt (e.g., we use user story instead of requirement document) when implementing \toolScrum. We follow a disordered communication pattern in the design of \toolScrum, because, in sprint meetings, every development role can provide their opinion (e.g., to simulate the planning poker process). Every development role, except the {\em Scrum Master}, reads the common context (e.g., description of the programming problem) from a common buffer. Then, every role provides a discussion comment and is saved back in the buffer. Therefore, every role is aware of all the comments. 
Then, the {\em Scrum Master} summarizes the entire discussion and derives a list of user stories for each development role. 
During the sprint, similarly to \water and \tdds, the four development roles carry out the development activities in sequence. At the end of the sprint, the agents will start another sprint meeting to discuss the next steps, such as releasing the code or needing to fix the code because of test failures.



\subsubsection{Self-Refinement}
We implement self-refinement~\cite{madaan2024self}, which tries to refine the LLM-generated result through iterative feedback, to further improve the generated artifacts from every development activity. In all three variations of \tool, we assign other agents to review the generated artifacts for every development activity and provide improvement suggestions. We assign the agents from both the downstream activity and the tester to examine the generated artifacts and provide suggestions. The suggestions are then considered for the re-generation of the artifacts. We include the tester in every development activity to emulate the DevTestOps practice~\cite{DevTestOps}, where testers are involved in all development activities and provide feedback on the quality aspects. For example, once a requirement engineer generates a requirement document, both the architect and tester would read the document and provide suggestions for improvement. Then, the requirement engineer will re-generate the requirement document based on the previously generated document and suggestions. 
At the development and testing activity, the tester will generate a test failure report if any of the LLM-generated tests fail or if the code cannot be executed (e.g., due to syntax error). The test failure report is then given to the developer for bug fixing. We repeat the process {\it t} times to self-refine the generated code. In our implementation, we currently set {\it t}=3. If the code still cannot pass the test, we repeat the entire software development process.

\subsection{Implementation and Experiment Settings}
\noindent\textbf{Environment.} We use GPT3.5 (version {\sf gpt-3.5-turbo-1106}) as our underlying LLM due to its popularity and wide usage in code generation research \cite{li2023structured,shinn2024reflexion,dong2023self}. We leverage \delete{the APIs provided by} OpenAI\add{'s APIs} \add{(version 0.28.1)} to interact with GPT. We send prompts using JSON format \cite{white2023chatgpt} and send all the conversation history as part of the prompt \cite{hong2023metagpt}. We set the temperature value to 0.8 and explore the effect of the temperature value in RQ3. We implemented \tool using \textit{Python 3.9}\add{.}\delete{and version \textit{0.28.1} for the OpenAI API.}

\noindent\textbf{Benchmark Datasets.} We follow prior studies \cite{zhou2023language,li2023structured,zhong2024ldb} and evaluate the code generation result using four benchmarks: HumanEval, HumanEval-ET, MBPP (Mostly Basic Python Programming), and MBPP-ET. These benchmarks contain both the programming problems and tests for evaluation. Given a programming problem, we consider that a generated code snippet is correct if it can pass all the provided tests. HumanEval~\cite{chen2021evaluating} has 164 programming problems, and  MBPP~\cite{austin2021program} has 427 programming problems \add{(we use the sanitized version released by the original authors)} and three test cases for each problem. We also use the dataset published by Dong et al.~\cite{dong2023codescore}, where they use the same problems as HumanEval and MBPP but offer stronger evaluation test cases (around 100 test cases for each problem, called HumanEval-ET and MBPP-ET). All these benchmarks use Python as the programming language. Each programming problem contains the INPUT pairs $<$method signature, method description, invoke examples$>$ and expect the code as OUTPUT.


\noindent\textbf{Evaluation Metric.} To evaluate the quality of the generated code, we use the Pass@K metric~\cite{dong2023self, chen2021evaluating}. Pass@K evaluates the first K generated code's functional accuracy (i.e., whether the generated code can pass all the test cases). In this work, we set K=1 to evaluate if the first generated code can pass all the provided test cases. A Pass@1 of 100 means 100\% of the generated code can pass all the tests in the first attempt. \add{We use Pass@1 because it is a stricter criterion, reflecting situations where developers do not have the groundtruth for automatically evaluating multiple attempts.}
\section{Results}\label{result}
\delete{In this section, }\add{We evaluate \tool with four research questions (RQs). }\delete{we present the research questions aimed at evaluating \tool. 
For each RQ, we discuss the motivation, the approach we take to answering it, and the results and discussion of the findings. }

\subsection*{RQ1: What is the code generation accuracy of \tool?}

\noindent{\underline{\textit{Motivation.}}}
In this RQ, we emulate the three process models using LLM agents and compare their results on code generations. Such results may provide invaluable evidence for future researchers seeking to optimize process models for code generation within their specific business domain.

\noindent{\underline{\textit{Approach.}}} 
As a baseline for comparison, we directly give the programming problems to ChatGPT (which we refer to as \raw). 
Although prior works~\cite{kang2023large,le2023log,nashid2023retrieval,li2023structured} show that few-shot learning can improve the results from LLMs, \add{they can be biased on how the few-shot samples are selected~\cite{xu2022alleviating}.}\delete{how the few-shot samples are selected can also introduce biases~\cite{xu2022alleviating}}. Hence, we use zero-shot learning in our experiment. To control for randomness in the experiment, we ensure all these experiments use the same temperature value (t=0.8) and the same model version (\textit{gpt-3.5-turbo-1106}). Finally, we repeat each \tool five times and 
report the average Pass@1 and standard deviation across the runs. We also conduct a student's t-test to study if \tool's results are statistically significantly different from \raw. 

\begin{table}
	\caption{Average and standard deviation of the Pass@1 accuracy across five runs, with the best Pass@1 marked in bold. The numbers in the parentheses show the percentage difference compared to \raw. Statistically significant differences are marked with a ``*''.}
	\label{tab:rq1_accuracy}
	\scriptsize
	\centering
 	\scalebox{0.85}{\setlength{\tabcolsep}{0.6mm}
	\begin{tabular}{lrrrr}
 
	\toprule	
 &HumanEval & HumanEval-ET & MBPP & MBPP-ET\\
	\midrule
	\raw & 64.4$\pm$3.7 & 49.8$\pm$3.0   & 77.5$\pm$0.8 & 53.9$\pm$0.7\\
                
	\toolWater & 69.5$\pm$2.3 (+7.9\%)* & 59.4$\pm$2.5 (+19.2\%)* & 76.3$\pm$0.9 (-1.5\%) & 51.1$\pm$1.7 (-5.2\%)* \\
	\toolTdd & 69.8$\pm$2.2 (+8.4\%)* & 60.0$\pm$2.1 (+20.5\%)* & 76.8$\pm$0.9 (-1.0\%) & 52.8$\pm$0.7 (-2.1\%)*  \\
	\toolScrum & {\bf 75.2$\pm$1.1 (+16.8\%)*} & {\bf 65.5$\pm$1.9 (+31.5\%)*} & {\bf 82.5$\pm$0.6 (+6.5\%)*} & {\bf 56.7$\pm$1.4 (+5.2\%)*} \\
	\bottomrule
	\end{tabular}
 	}
  \vspace{-3mm}

\end{table}

\noindent{\underline{\textit{Result.}}} \textit{\textbf{\toolScrum shows a consistent improvement over \raw, achieving 5.2\% to 31.5\% improvement in Pass@1. }}
Table~\ref{tab:rq1_accuracy} shows the Pass@1 accuracy of \tool across different process models on the benchmark datasets studied. 
As shown in the Table~\ref{tab:rq1_accuracy}, for HumanEval and HumanEval-ET, all of the studied process models have 7.9\% to even 31.5\% improvement in Pass@1 compared to \raw, and the improvements are all statistically significant (p $<=$ 0.05). 
For MBPP and MBPP-ET, \toolScrum also has statistically significant improvements of 5.2\% to 6.5\%, even though we see a slight decrease when adopting \toolWater and \toolTdd to MBPP and MBPP-ET.

Despite slight variations in code generation responses form LLM across executions, \textbf{we find stable standard deviations of Pass@1, ranging from 0.6\% to 3.7\% across all process models and benchmarks}. In particular, \toolScrum has the lowest standard deviation (0.6\% to 1.9\%, an average of 1.2\%), while \raw has the highest standard deviation (0.5\% to 3.7\%, an average of 2\%). Following \toolScrum, \toolWater has the second highest standard deviation, with \toolTdd is ranking third. In conclusion, although the models generally have consistent Pass@1 across runs, \toolScrum consistently produces the most stable results. 

\begin{table}
    \centering
    \caption{\tool test failure categorization. Failure types are generated from Python Interpreter~\cite{pythonInterpreter}. Darker red indicates higher percentages of the failure categories in the generated code across the models. Percentages are calculated by the ratio of specific failure types to the total number of failed tests across different process models.}
    \scalebox{0.65}{\setlength{\tabcolsep}{0.3mm}
\begin{tabular}{ll|llllllll|c}
\toprule
     &        &                \multicolumn{8}{c|}{\textbf{Failure  Categories}} &                            \\
   Benchmarks  &       Model &                     Assertion &                         Syntax &                         Name &                         Type &                       Index &                        Value &                     Recursion &                     Attribute &                           Total \\
\midrule
    HumanEval &      RawGPT &   \cellcolor{red!24}36 (24\%) &  \cellcolor{red!100}18 (100\%) &  \cellcolor{red!58}11 (58\%) &   \cellcolor{red!17}2 (17\%) &  \cellcolor{red!50}2 (50\%) &   \cellcolor{red!12}1 (12\%) &      \cellcolor{red!0}0 (0\%) &      \cellcolor{red!0}0 (0\%) &   \cellcolor{red!33}70 (33\%) \\
     &   Waterfall &   \cellcolor{red!26}39 (26\%) &       \cellcolor{red!0}0 (0\%) &   \cellcolor{red!16}3 (16\%) &   \cellcolor{red!25}3 (25\%) &    \cellcolor{red!0}0 (0\%) &   \cellcolor{red!12}1 (12\%) &      \cellcolor{red!0}0 (0\%) &      \cellcolor{red!0}0 (0\%) &   \cellcolor{red!22}46 (22\%) \\
     &  TDD &   \cellcolor{red!26}39 (26\%) &       \cellcolor{red!0}0 (0\%) &   \cellcolor{red!21}4 (21\%) &   \cellcolor{red!25}3 (25\%) &  \cellcolor{red!25}1 (25\%) &   \cellcolor{red!62}5 (62\%) &      \cellcolor{red!0}0 (0\%) &      \cellcolor{red!0}0 (0\%) &   \cellcolor{red!24}52 (24\%) \\
     &       Scrum &   \cellcolor{red!25}38 (25\%) &       \cellcolor{red!0}0 (0\%) &     \cellcolor{red!5}1 (5\%) &   \cellcolor{red!33}4 (33\%) &  \cellcolor{red!25}1 (25\%) &   \cellcolor{red!12}1 (12\%) &      \cellcolor{red!0}0 (0\%) &      \cellcolor{red!0}0 (0\%) &   \cellcolor{red!21}45 (21\%) \\
     \midrule
 HumanEval-ET &      RawGPT &   \cellcolor{red!21}39 (21\%) &     \cellcolor{red!82}9 (82\%) &  \cellcolor{red!43}13 (43\%) &   \cellcolor{red!25}1 (25\%) &  \cellcolor{red!57}4 (57\%) &   \cellcolor{red!18}2 (18\%) &      \cellcolor{red!0}0 (0\%) &      \cellcolor{red!0}0 (0\%) &   \cellcolor{red!27}68 (27\%) \\ 
 &   Waterfall &   \cellcolor{red!26}49 (26\%) &       \cellcolor{red!9}1 (9\%) &   \cellcolor{red!23}7 (23\%) &     \cellcolor{red!0}0 (0\%) &  \cellcolor{red!14}1 (14\%) &   \cellcolor{red!27}3 (27\%) &      \cellcolor{red!0}0 (0\%) &      \cellcolor{red!0}0 (0\%) &   \cellcolor{red!24}61 (24\%) \\
  &  TDD &   \cellcolor{red!26}50 (26\%) &       \cellcolor{red!9}1 (9\%) &   \cellcolor{red!20}6 (20\%) &   \cellcolor{red!50}2 (50\%) &  \cellcolor{red!14}1 (14\%) &   \cellcolor{red!36}4 (36\%) &      \cellcolor{red!0}0 (0\%) &      \cellcolor{red!0}0 (0\%) &   \cellcolor{red!25}64 (25\%) \\
 &       Scrum &   \cellcolor{red!27}52 (27\%) &       \cellcolor{red!0}0 (0\%) &   \cellcolor{red!13}4 (13\%) &   \cellcolor{red!25}1 (25\%) &  \cellcolor{red!14}1 (14\%) &   \cellcolor{red!18}2 (18\%) &      \cellcolor{red!0}0 (0\%) &      \cellcolor{red!0}0 (0\%) &   \cellcolor{red!24}60 (24\%) \\
  \midrule
         MBPP &      RawGPT &   \cellcolor{red!23}66 (23\%) &     \cellcolor{red!70}7 (70\%) &   \cellcolor{red!41}7 (41\%) &   \cellcolor{red!25}7 (25\%) &  \cellcolor{red!67}2 (67\%) &     \cellcolor{red!0}0 (0\%) &  \cellcolor{red!100}1 (100\%) &    \cellcolor{red!50}1 (50\%) &   \cellcolor{red!26}91 (26\%) \\
          &   Waterfall &   \cellcolor{red!27}76 (27\%) &     \cellcolor{red!10}1 (10\%) &   \cellcolor{red!29}5 (29\%) &   \cellcolor{red!21}6 (21\%) &    \cellcolor{red!0}0 (0\%) &     \cellcolor{red!0}0 (0\%) &      \cellcolor{red!0}0 (0\%) &      \cellcolor{red!0}0 (0\%) &   \cellcolor{red!25}88 (25\%) \\
          &  TDD &   \cellcolor{red!30}86 (30\%) &     \cellcolor{red!10}1 (10\%) &   \cellcolor{red!29}5 (29\%) &   \cellcolor{red!29}8 (29\%) &    \cellcolor{red!0}0 (0\%) &     \cellcolor{red!0}0 (0\%) &      \cellcolor{red!0}0 (0\%) &      \cellcolor{red!0}0 (0\%) &  \cellcolor{red!29}100 (29\%) \\
          &       Scrum &   \cellcolor{red!20}58 (20\%) &     \cellcolor{red!10}1 (10\%) &     \cellcolor{red!0}0 (0\%) &   \cellcolor{red!25}7 (25\%) &  \cellcolor{red!33}1 (33\%) &     \cellcolor{red!0}0 (0\%) &      \cellcolor{red!0}0 (0\%) &    \cellcolor{red!50}1 (50\%) &   \cellcolor{red!20}68 (20\%) \\
           \midrule
      MBPP-ET &      RawGPT &  \cellcolor{red!24}133 (24\%) &     \cellcolor{red!57}8 (57\%) &  \cellcolor{red!26}22 (26\%) &  \cellcolor{red!24}26 (24\%) &  \cellcolor{red!50}2 (50\%) &     \cellcolor{red!8}3 (8\%) &  \cellcolor{red!100}1 (100\%) &  \cellcolor{red!100}1 (100\%) &  \cellcolor{red!24}196 (24\%) \\
       &   Waterfall &  \cellcolor{red!27}148 (27\%) &     \cellcolor{red!14}2 (14\%) &  \cellcolor{red!26}22 (26\%) &  \cellcolor{red!25}27 (25\%) &    \cellcolor{red!0}0 (0\%) &  \cellcolor{red!39}15 (39\%) &      \cellcolor{red!0}0 (0\%) &      \cellcolor{red!0}0 (0\%) &  \cellcolor{red!27}214 (27\%) \\
       &  TDD &  \cellcolor{red!27}147 (27\%) &     \cellcolor{red!14}2 (14\%) &  \cellcolor{red!27}23 (27\%) &  \cellcolor{red!26}28 (26\%) &  \cellcolor{red!25}1 (25\%) &  \cellcolor{red!26}10 (26\%) &      \cellcolor{red!0}0 (0\%) &      \cellcolor{red!0}0 (0\%) &  \cellcolor{red!26}211 (26\%) \\
       &       Scrum &  \cellcolor{red!22}123 (22\%) &     \cellcolor{red!14}2 (14\%) &  \cellcolor{red!21}18 (21\%) &  \cellcolor{red!26}28 (26\%) &  \cellcolor{red!25}1 (25\%) &  \cellcolor{red!26}10 (26\%) &      \cellcolor{red!0}0 (0\%) &      \cellcolor{red!0}0 (0\%) &  \cellcolor{red!23}182 (23\%) \\
       \bottomrule
\end{tabular}
}
\vspace{-3mm}
    \label{tab:RQ1_Failure}
\end{table}

\noindent\textit{\textbf{There are potential issues in the tests provided by the benchmarks, which may hinder the Pass@1 of \tool.}} Table~\ref{tab:RQ1_Failure} provides a breakdown of failure types from the Python Interpreter~\cite{pythonInterpreter} 
 across various process models and benchmarks. \add{For example, \texttt{IndexError} happens when the generated code does not handle an out-of-bound index, causing an exception to be thrown.} While we repeat our experiment 5 times, the standard deviation across runs is low; hence, we represent test failure from only one of the runs. Aligned with the findings from Table~\ref{tab:rq1_accuracy}, \toolScrum has the lowest \texttt{AssertionError} compared to other models \add{(i.e., higher pass rate)}. We also notice that \texttt{SyntaxError} is more evident in \raw\delete{across all benchmarks}, as expected due to the absence of a code review and testing process. However, there are still higher test failures in \toolWater and \toolTdd caused by increased occurrences of \texttt{ValueError}, \texttt{TypeError}, \add{\texttt{IndexError}} and \texttt{NameError}, for MBPP and MBPP-ET as seen in{\delete{Table~\ref{tab:rq1_accuracy}}} {\add{Table~\ref{tab:RQ1_Failure}}}.  
Upon manual investigation of \add{prevalent} test failures types, we discover that \toolWater and \toolTdd~{\textbf{introduce various validations and enforce programming naming conventions in the generated code,}} \textit{\textbf{which may help improve code quality but cause tests to fail}}. For example, Listing~\ref{lst:MBPP-582} depicts the provided tests and the generated code{\delete{/test}} for MBPP-582, of which the objective is to \textit{``Write a function to check if a dictionary is empty or not''}. While \raw passed the provided tests, \toolWater and \toolTdd failed. This failure is because the generated code contains {\delete{validation}} {\add{\textit{strict input validation}}} to check that the input should be of type \textit{dict}. However, the MBPP-582 provided test uses an input of the type \textit{set}, which fails the validation, causing a \texttt{TypeError} exception. Moreover, \toolWater and \toolTdd enforce the common naming convention format and more meaningful function name (e.g., \texttt{my\_dict} v.s. \texttt{is\_dict\_empty}), both of which causes \texttt{NameError} exception due to {\delete{missing declaration}}{\add{\textit{wrong function declaration}}}, causing test failure. More interestingly, we find that such code standardization may also be misled by the requirement provided by the benchmark itself. For example, the MBPP-582 requirement specifies expected input as \texttt{dict}, yet provides a \texttt{set} type as the input to the test. The LLM code generation indeed captures this correct requirement by validating that input must be of type \texttt{dict}. Such inconsistency in the benchmark may reduce the Pass@1. 

\begin{lstlisting}[language=PYTHON,caption={MBPP-582 Test Failure due to \textit{Strict Input Validation} and  \textit{Wrong Function Name}.  },label={lst:MBPP-582}, basicstyle=\scriptsize\ttfamily]
# MBPP-582: check if a dictionary is empty
# MBPP Test Case
def Test(): 
    assert my_dict({10})==False # {10} is a set not a dict
# rawGPT's answer
def my_dict(dict1): 
    return len(dict1) == 0
# Waterfall/TDD's answer, enforces the input type must be a dictionary
def is_dict_empty(input_dict): # function name is renamed from my_dict
    if not isinstance(input_dict, dict): 
        raise TypeError("Input is not a dictionary")
    return True if not input_dict else False
\end{lstlisting}

In MBPP-794 (Listing~\ref{lst:MBPP-794_797}), test cases provided by MBPP-ET change the return value from a {\tt boolean} (as is the case in MBPP) to the string {\tt ``match!''}. Moreover, in MBPP-797, MBPP-ET’s test capitalized the last word in uppercase ({\tt range} v.s. {\tt ``R''ange}). Such non-standard evaluation leads to unfair test results (leads to failure), which may bias the experimental results for MBPP-ET. Such bias suggests that the decrease in Pass@1 rates for MBPP-ET is not solely due to an increase in the number of provided tests. 

\begin{lstlisting}[language=PYTHON,caption={MBPP-794 \& MBPP-797 Test Failure due to Irregularity in Test Cases.},label={lst:MBPP-794_797}, basicstyle=\scriptsize\ttfamily]
# Example1: Changed return type from boolean to string
assert text_starta_endb("aabbbb") # MBPP 794
assert text_starta_endb("aabbbb")==('match!') # MBPP-ET 794
# Example2: Capitalized the last character in function name
assert sum_in_range(2,5) == 8 # MBPP-797
assert sum_in_Range(2,5) == 8 # MBPP-ET-797
\end{lstlisting}

Fixing the issues largely improve \tool's Pass@1: 16 to 28 improvement in HumanEval, 29 to 35 in HumanEval-ET, 15 to 21 in MBPP, and 28 to 42 in MBPP-ET. Namely, \tool's Pass@1 can achieve over 90 to 95 across all benchmarks. Even though we also observe improvement in \raw's Pass@1, the three \tool models had greater improvement. On average, the three models have a Pass@1 that is 14\%, 8\%, 1.4\%, and 5\% better than \raw in HumanEval, HumanEval-ET, MBPP, and MBPP-ET, respectively. 

The \delete{aggregated}results underscore the deficiencies in the benchmarks, suggesting that the current Pass@1 score of \tool \delete{(Table~\ref{tab:rq1_accuracy}) }could represent a lower bound. These preliminary findings highlight the potential of \tool and suggest that future research should improve the benchmarks by incorporating input checking or consistent naming convention into the tests and subsequently re-evaluate existing code generation techniques.  

\rqboxc{\toolScrum achieves the best results, with a Pass@1 that is 5.2\% to 31.5\% better than \raw. \toolScrum also has the most stable results (average standard decision of 1.3\% across all benchmarks) among all models. Notably, while \toolWater and \toolTdd enhance code quality, such improvements may result in test failures. }

\subsection*{RQ2: How do different development activities impact the quality of the generated code?}

\begin{table*}
\centering
    \caption{Pass@1 and \textit{Error/Warning/Convention/Refactor/Handled-Exception} density (per 10 lines of code) in the full \tool (with all the development activities) and after removing a development activity. A \textit{\textbf{lower}} error/warning/convention/refactor is preferred, and a \textit{\textbf{higher}} handled-exception is preferred. Darker red
indicates a larger decrease in percentages, while darker green indicates a larger increase in percentages. } 
    \label{tab:smell}
    \setlength{\tabcolsep}{0.25mm}
    	\scalebox{0.7}{
\begin{tabular}{llrrrrrr|rrrrrr}
\toprule
\multirow{3}{*}{ Model } & \multirow{3}{*}{Dev. Activities  } & \multicolumn{6}{c}{HumanEval} & \multicolumn{6}{c}{MBPP} \\ \cline{3-8}\cline{9-14}\rule{0pt}{2.3ex}
& & \multirow{2}{*}{ Pass@1 } & \multirow{2}{*}{ Error } & \multirow{2}{*}{ Warning } & \multirow{2}{*}{ Convention } & \multirow{2}{*}{ Refactor } & Handled & \multirow{2}{*}{ Pass@1 } & \multirow{2}{*}{ Error } & \multirow{2}{*}{ Warning } & \multirow{2}{*}{ Convention } & \multirow{2}{*}{ Refactor } &  Handled  \\
& & & & & & & Exception  & & & & & & Exception  \\
\midrule
\raw & -- & 64.4 & 0.25 & 0.19 & 0.39 & 0.30 & 0.00 & 77.47 & 0.22 & 0.20 & 1.18 & 0.30 & 0.01 \\
\midrule
\multirow{5}{*}{ \toolWater } & full & 69.5 & 0.01 & 0.12 & 0.24 & 0.21 & 0.37 & 76.35 & 0.03 & 0.12 & 0.47 & 0.23 & 0.67 \\

 &   rm-requirement &    \cellcolor{red!1.7} -1.2 (1.7\%) &                              0.0 (0.0\%) &     \cellcolor{red!8.3} -0.01 (8.3\%) &     \cellcolor{red!8.3} -0.02 (8.3\%) &     \cellcolor{red!4.8} -0.01 (4.8\%) &   \cellcolor{red!16.2} -0.06 (16.2\%) &  \cellcolor{green!0.9} +0.7 (0.9\%) &     \cellcolor{red!66.7} -0.02 (66.7\%) &     \cellcolor{red!16.7} -0.02 (16.7\%) &                           0.0 (0.0\%) &   \cellcolor{green!8.7} +0.02 (8.7\%) &   \cellcolor{red!13.4} -0.09 (13.4\%) \\
 &        rm-design &    \cellcolor{red!1.7} -1.2 (1.7\%) &  \cellcolor{green!100.0} +0.01 (100.0\%) & \cellcolor{green!16.7} +0.02 (16.7\%) &   \cellcolor{green!8.3} +0.02 (8.3\%) &                           0.0 (0.0\%) &   \cellcolor{red!40.5} -0.15 (40.5\%) &   \cellcolor{red!2.1} -1.64 (2.1\%) &     \cellcolor{red!33.3} -0.01 (33.3\%) &                             0.0 (0.0\%) &   \cellcolor{green!4.3} +0.02 (4.3\%) &   \cellcolor{green!4.3} +0.01 (4.3\%) &    \cellcolor{red!29.9} -0.2 (29.9\%) \\
 &    rm-codeReview &    \cellcolor{red!3.5} -2.4 (3.5\%) &                              0.0 (0.0\%) &                           0.0 (0.0\%) &   \cellcolor{red!12.5} -0.03 (12.5\%) &   \cellcolor{green!9.5} +0.02 (9.5\%) &   \cellcolor{red!24.3} -0.09 (24.3\%) & \cellcolor{green!0.6} +0.46 (0.6\%) &     \cellcolor{red!66.7} -0.02 (66.7\%) &   \cellcolor{green!58.3} +0.07 (58.3\%) &     \cellcolor{red!4.3} -0.02 (4.3\%) &     \cellcolor{red!8.7} -0.02 (8.7\%) &   \cellcolor{red!25.4} -0.17 (25.4\%) \\
 &          rm-test & \cellcolor{red!56.1} -39.0 (56.1\%) & \cellcolor{green!100.0} +0.17 (1700.0\%) &   \cellcolor{green!8.3} +0.01 (8.3\%) & \cellcolor{green!29.2} +0.07 (29.2\%) &  \cellcolor{green!47.6} +0.1 (47.6\%) &  \cellcolor{green!27.0} +0.1 (27.0\%) & \cellcolor{red!31.0} -23.7 (31.0\%) & \cellcolor{green!100.0} +0.09 (300.0\%) &   \cellcolor{green!41.7} +0.05 (41.7\%) & \cellcolor{green!76.6} +0.36 (76.6\%) &   \cellcolor{green!4.3} +0.01 (4.3\%) &     \cellcolor{red!1.5} -0.01 (1.5\%) \\

\midrule
\multirow{5}{*}{ \toolTdd } & full  & 69.8 & 0.01 & 0.08 & 0.33 & 0.27 & 0.33 & 76.77 & 0.04 & 0.13 & 0.71 & 0.28 & 0.62 \\

 &   rm-requirement &    \cellcolor{red!4.2} -2.9 (4.2\%) &  \cellcolor{green!100.0} +0.01 (100.0\%) & \cellcolor{green!12.5} +0.01 (12.5\%) &                           0.0 (0.0\%) &   \cellcolor{red!11.1} -0.03 (11.1\%) &    \cellcolor{red!30.3} -0.1 (30.3\%) & \cellcolor{green!2.5} +1.92 (2.5\%) &     \cellcolor{red!50.0} -0.02 (50.0\%) &   \cellcolor{green!15.4} +0.02 (15.4\%) &   \cellcolor{green!4.2} +0.03 (4.2\%) &                           0.0 (0.0\%) &   \cellcolor{red!43.5} -0.27 (43.5\%) \\
 &        rm-design &    \cellcolor{red!4.2} -2.9 (4.2\%) &                              0.0 (0.0\%) & \cellcolor{green!25.0} +0.02 (25.0\%) &   \cellcolor{red!18.2} -0.06 (18.2\%) & \cellcolor{green!11.1} +0.03 (11.1\%) &   \cellcolor{red!39.4} -0.13 (39.4\%) & \cellcolor{green!1.6} +1.22 (1.6\%) &     \cellcolor{red!50.0} -0.02 (50.0\%) &     \cellcolor{red!23.1} -0.03 (23.1\%) &     \cellcolor{red!4.2} -0.03 (4.2\%) & \cellcolor{green!14.3} +0.04 (14.3\%) &   \cellcolor{red!38.7} -0.24 (38.7\%) \\
 &    rm-codeReview &    \cellcolor{red!0.7} -0.5 (0.7\%) &    \cellcolor{red!100.0} -0.01 (100.0\%) & \cellcolor{green!25.0} +0.02 (25.0\%) &   \cellcolor{red!12.1} -0.04 (12.1\%) & \cellcolor{green!11.1} +0.03 (11.1\%) &   \cellcolor{red!27.3} -0.09 (27.3\%) & \cellcolor{green!1.3} +0.98 (1.3\%) &     \cellcolor{red!75.0} -0.03 (75.0\%) &     \cellcolor{green!7.7} +0.01 (7.7\%) &     \cellcolor{red!7.0} -0.05 (7.0\%) &  \cellcolor{green!35.7} +0.1 (35.7\%) &   \cellcolor{red!14.5} -0.09 (14.5\%) \\
 &          rm-test & \cellcolor{red!17.0} -11.9 (17.0\%) &  \cellcolor{green!100.0} +0.07 (700.0\%) & \cellcolor{green!50.0} +0.04 (50.0\%) &     \cellcolor{red!6.1} -0.02 (6.1\%) &   \cellcolor{red!18.5} -0.05 (18.5\%) &    \cellcolor{red!30.3} -0.1 (30.3\%) & \cellcolor{red!22.5} -17.3 (22.5\%) & \cellcolor{green!100.0} +0.11 (275.0\%) & \cellcolor{green!100.0} +0.14 (107.7\%) & \cellcolor{green!22.5} +0.16 (22.5\%) &     \cellcolor{red!3.6} -0.01 (3.6\%) & \cellcolor{green!25.8} +0.16 (25.8\%) \\

\midrule
\multirow{7}{*}{ \toolScrum } & full & 75.2 & 0.00 & 0.13 & 0.21 & 0.24 & 0.15 & 82.48 & 0.02 & 0.17 & 0.51 & 0.23 & 0.44 \\

 &   rm-requirement &  \cellcolor{green!1.3} +1.0 (1.3\%) &                              0.0 (0.0\%) &                           0.0 (0.0\%) & \cellcolor{green!23.8} +0.05 (23.8\%) &     \cellcolor{red!4.2} -0.01 (4.2\%) & \cellcolor{green!60.0} +0.09 (60.0\%) &   \cellcolor{red!2.3} -1.92 (2.3\%) &   \cellcolor{green!50.0} +0.01 (50.0\%) &                             0.0 (0.0\%) &   \cellcolor{green!2.0} +0.01 (2.0\%) &     \cellcolor{red!4.3} -0.01 (4.3\%) &   \cellcolor{green!9.1} +0.04 (9.1\%) \\
 &        rm-design &  \cellcolor{green!1.3} +1.0 (1.3\%) &                              0.0 (0.0\%) &   \cellcolor{red!15.4} -0.02 (15.4\%) &   \cellcolor{red!14.3} -0.03 (14.3\%) &                           0.0 (0.0\%) &   \cellcolor{red!20.0} -0.03 (20.0\%) & \cellcolor{green!1.1} +0.89 (1.1\%) &     \cellcolor{red!50.0} -0.01 (50.0\%) &     \cellcolor{red!23.5} -0.04 (23.5\%) & \cellcolor{green!21.6} +0.11 (21.6\%) & \cellcolor{green!13.0} +0.03 (13.0\%) &   \cellcolor{red!43.2} -0.19 (43.2\%) \\
 &    rm-codeReview &    \cellcolor{red!2.7} -2.0 (2.7\%) &                              0.0 (0.0\%) &                           0.0 (0.0\%) &   \cellcolor{red!14.3} -0.03 (14.3\%) &                           0.0 (0.0\%) &   \cellcolor{red!20.0} -0.03 (20.0\%) & \cellcolor{green!0.2} +0.19 (0.2\%) &                             0.0 (0.0\%) &       \cellcolor{red!5.9} -0.01 (5.9\%) &   \cellcolor{green!2.0} +0.01 (2.0\%) & \cellcolor{green!26.1} +0.06 (26.1\%) &    \cellcolor{red!22.7} -0.1 (22.7\%) \\
 &          rm-test & \cellcolor{red!18.9} -14.2 (18.9\%) &  \cellcolor{green!100.0} +0.03 (588.2\%) & \cellcolor{green!46.2} +0.06 (46.2\%) &                           0.0 (0.0\%) &   \cellcolor{red!12.5} -0.03 (12.5\%) & \cellcolor{green!46.7} +0.07 (46.7\%) & \cellcolor{red!32.1} -26.5 (32.1\%) & \cellcolor{green!100.0} +0.13 (650.0\%) &    \cellcolor{green!58.8} +0.1 (58.8\%) & \cellcolor{green!80.4} +0.41 (80.4\%) &   \cellcolor{red!13.0} -0.03 (13.0\%) & \cellcolor{green!31.8} +0.14 (31.8\%) \\
 & rm-sprintMeeting &    \cellcolor{red!2.1} -1.6 (2.1\%) &                              0.0 (0.0\%) &     \cellcolor{red!7.7} -0.01 (7.7\%) &     \cellcolor{red!9.5} -0.02 (9.5\%) &   \cellcolor{red!12.5} -0.03 (12.5\%) &  \cellcolor{green!66.7} +0.1 (66.7\%) &   \cellcolor{red!3.7} -3.09 (3.7\%) &                             0.0 (0.0\%) &     \cellcolor{red!11.8} -0.02 (11.8\%) &   \cellcolor{green!9.8} +0.05 (9.8\%) &     \cellcolor{red!4.3} -0.01 (4.3\%) & \cellcolor{green!47.7} +0.21 (47.7\%) \\
\bottomrule
\end{tabular}
}
\vspace{-3mm}

\end{table*}

\noindent{\underline{\textit{Motivation.}}}
As observed in RQ1, various process models can indeed affect the functional correctness (Pass@1) of the generated code. However, it is equally crucial to understand code quality issues such as code smells and the impact of a development activity on the generated code. This understanding is essential for assessing whether the generated code adheres to industry best practices. Moreover, such insight may offer valuable opportunities for enhancing the design, readability, and maintainability in auto-generated code. 



\noindent{\underline{\textit{Approach.}}}  
To study the impact of each development activity on code quality, we remove each activity separately and re-execute \tool. For example, we first remove the requirement activity in \toolWater and execute \toolWater. Then, we add the requirement activity back and remove the design activity. We repeat the same process for every development activity. Note that we cannot remove the coding activity since our goal is code generation. Hence, we removed code review at the end of the coding activity. 

Code quality considers numerous facets beyond mere functional correctness~\cite{yeticstiren2023evaluating,yamashita2012code}. Other factors, such as code smells, maintainability, and readability, are also related to code quality. Hence, to gain a comprehensive understanding of how code quality changes, \delete{in addition to Pass@1,} we 1) apply a static code analyzer to detect code smells in the generated code and 2) study code reliability by analyzing the exception handling code. 
To study code smell, design, and readability, we apply Pylint 3.0.4~\cite{pylint,dasgupta2017code} (a Python static code analyzer) on the generated code. Pylint classifies the detected code smells into different categories such as {\em error}, {\em warning}, \textit{convention}, and \textit{refactor}. 

We study how the number of detected code smells in each category changes when removing an activity. 
Since the generated code may have different lengths, we report the density of the code smells in each category. We calculate the code smell density as the total number of code smell instances in a category (e.g., \textit{error}) divided by the total lines of code. 
To study reliability, we calculate the density of handled exceptions (total number of handled exceptions divided by the total lines of code) since exceptions are one of the most important mechanisms to detect and recover from errors~\cite{fetzer2004automatic}. For better visualization, we present the density results as per every 10 lines of code. We also ensure reliability of our results by repeating all of the aforementioned approach three times.

\noindent{\underline{\textit{Result.}}} 
\noindent\textbf{\textit{Testing has the largest impact on the functional correctness of the code, while other development activities only have small impacts. Removing testing causes Pass@1 to decrease by 17.0\% to 56.1\%. }}
Table~\ref{tab:smell} presents changes in Pass@1 and the densities of code smell and handled exceptions. We show the results for HumanEval and MBPP because they share the same programming problems and generated code with the other two benchmarks. 
Among all development activities, testing has the largest impact on Pass@1, where removing testing causes a large decrease in Pass@1 (17.0\% to 56.1\% decrease). The finding implies that LLM's generated tests are effective in improving the functional correctness of code. In both benchmarks, removing sprint meetings in \toolScrum also causes Pass@1 to drop. However, removing other activities only has a small and inconsistent effect on Pass@1. 
For example, in HumanEval, removing requirement, design, and code review generally causes Pass@1 to decrease (except for \toolScrum), but removing these activities improves Pass@1 in MBPP. 
In other words, most development activities do not significantly contribute to the functional correctness of the generated code. 

\delete{We further study how removing different activities impacts the code smell densities.}As shown in Table~\ref{tab:smell}, 
eliminating test activities significantly boosts \textit{error} and \textit{warning} smell densities by an average of 702.2\% and 52.2\%, respectively. Omitting design raises refactor smell density by an average of 7.1\%, and skipping code review leads to a 14.2\% average increase in \textit{warning} density. However, removing other development activities shows either a small or inconsistent impact. 
\add{We also find some differences in the artifacts generated by different roles. For example, although both roles generate documents, requirement engineers specify the acceptance criteria, while architects address time/space complexity.} 
In short, the findings show that {\textbf{adding design, testing, and code review can help reduce the density of code smell in the generated code}}.



\noindent\textbf{\textit{Having design and code review activities significantly improves reliability by increasing the density of handled exceptions, while other development activities only have small or no impacts.}} Removing design and code review activities separately causes the handled exception density to decrease from 20.0\% to 43.2\% and 14.5\% to 27.3\%, respectively. Namely, \delete{design and code review}\add{these two} activities add exception handling in the generated code, which may help improve reliability. 
Removing other \delete{development}activities shows a mixed relationship with the density of the handled exception. For example, removing testing in \toolTdd causes an increase of handled exception density by 25.8\% in MBPP (i.e., testing removes exception handling code) while causing a decrease of 30.3\% in HumanEval (i.e., testing adds exception handling code). While the effect of each development may be related to the nature of the benchmarks, 
our findings show that, in both benchmarks, \textbf{adding design and code review activities can help improve code reliability by handling more exceptions in the generated code}. 

\noindent\textbf{\textit{\tool shows consistent improvement over \raw in the quality of the generated code: decreasing the density of error/warning/convention/refactor code smells (6.7\% to 96.0\%) while significantly increasing handled exception density.}}
The code generated by \raw has higher \textit{error/warning/convention/refactor} code smell densities than that of \toolWater, \toolTdd, and \toolScrum. This finding shows all three models improve the quality of the generated code to different degrees. 
Specifically, compared to \raw, \tool decreases the 
\textit{error} code smell density by 81.8\% to 96.0\%, \textit{warning} density by 15.0\% to 57.9\%, \textit{convention} density by 15.4\% to 60.2\%, and \textit{refactor} density by 6.7\% to 30.0\%.
Meanwhile, \raw has fewer handled exceptions than \tool. As Table~\ref{tab:smell} shows, in both HumanEval and MBPP, \raw has almost zero handled exception, while \toolWater generates the most handled-exception (0.37 and 0.67 handled exceptions per every 10 LOC in the two benchmarks), \toolTdd ranks second (0.33 and 0.62), and then \toolScrum (0.15 and 0.44). 
\delete{In conclusion, these findings show that }\add{In short, }\tool improves the quality of the generated code by reducing code smells while adding more exception-handling code. 

\rqboxc{Compared to \raw, \tool remarkably improves the quality of the generated code by reducing code smells and adding more exception handling. Testing has the most significant impact on Pass@1 and code smells among all development activities, while having design and code review greatly improve the exception-handling ability. }
\subsection*{RQ3: How stable is the \tool generated code?}

\noindent{\underline{\textit{Motivation.}}}
In LLM, the stability of generated responses can be influenced by several parameters: 1) \textit{temperature}, affecting the randomness in the generated responses, and 2) \textit{model versions}, which may introduce variability due to changes in optimization and fine-tuning~\cite{Chen2023HowIC}. Understanding and improving the stability of LLMs is crucial for enhancing their trustworthiness, thereby facilitating their adoption in practice. Therefore, in this research question (RQ), we investigate the stability of our \tool in Pass@1 across four benchmarks, considering various temperature values and model versions.

\noindent{\underline{\textit{Approach.}}} 
We evaluate the Pass@1 of \tool across four versions of GPT3.5: \textit{turbo-0301}, \textit{turbo-0613}, \textit{turbo-1106}, and \textit{turbo-0125}. The latest version is \textit{turbo-0125} (published in January 2024), and the earlier version is \textit{turbo-0301} (published in March 2023). 
To avoid the effect of the model version when we vary the temperature, we use the same model version (\textit{turbo-1106}, the version that we used in prior RQs) to study the effect of temperature values. We set the temperature to 0.2, 0.4, 0.6, and 0.8 in our experiment. 
We execute \raw and the three variants of \tool three times under each configuration and report the average Pass@1.


\begin{figure}

  \centering
  \includegraphics[width=0.5\textwidth]{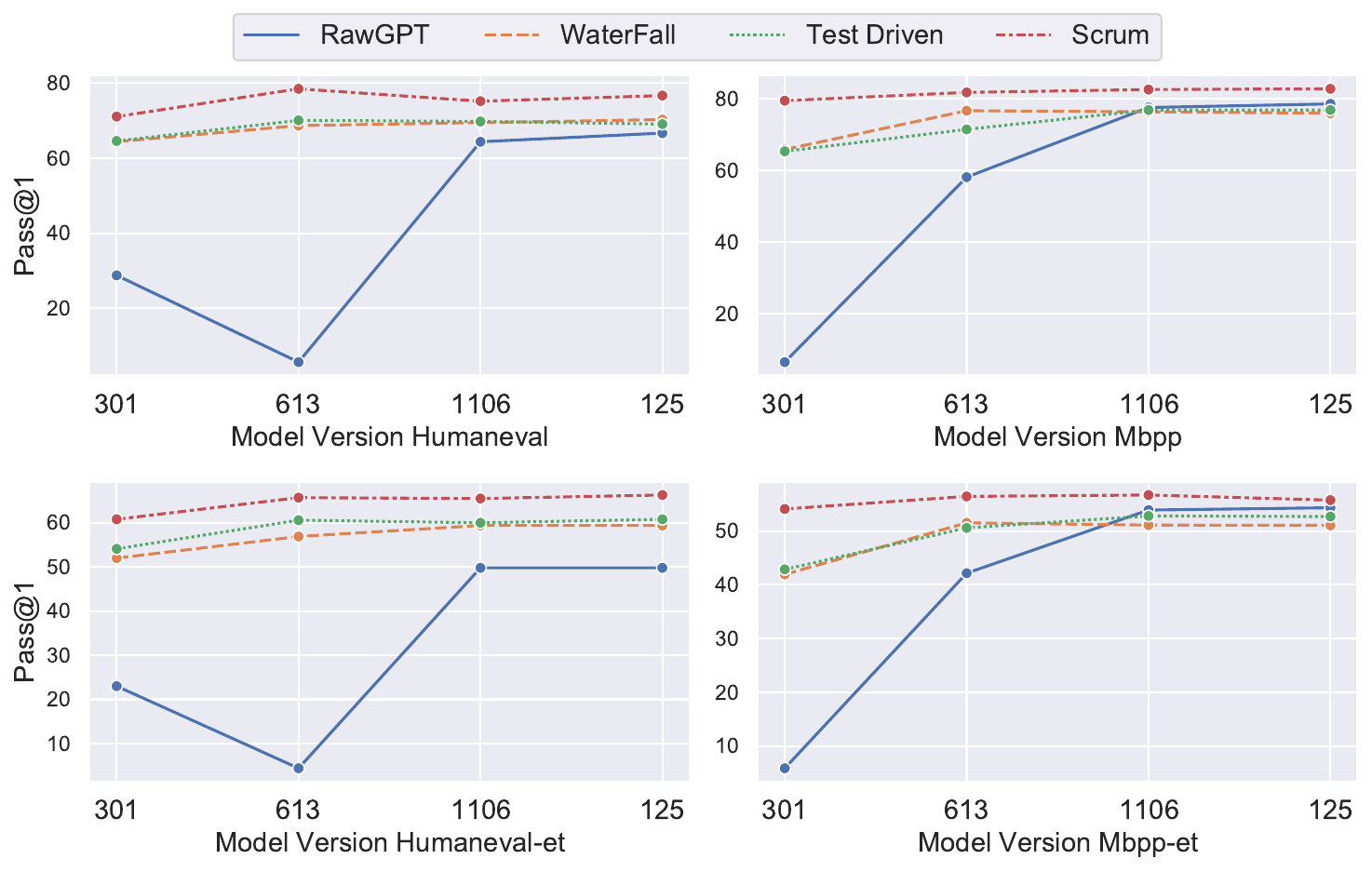}
  \caption{Pass@1 across GPT3.5 versions. }\label{fig:version}
  \vspace{-4mm}
\end{figure}

\begin{figure}
  \centering
  \includegraphics[width=0.5\textwidth]{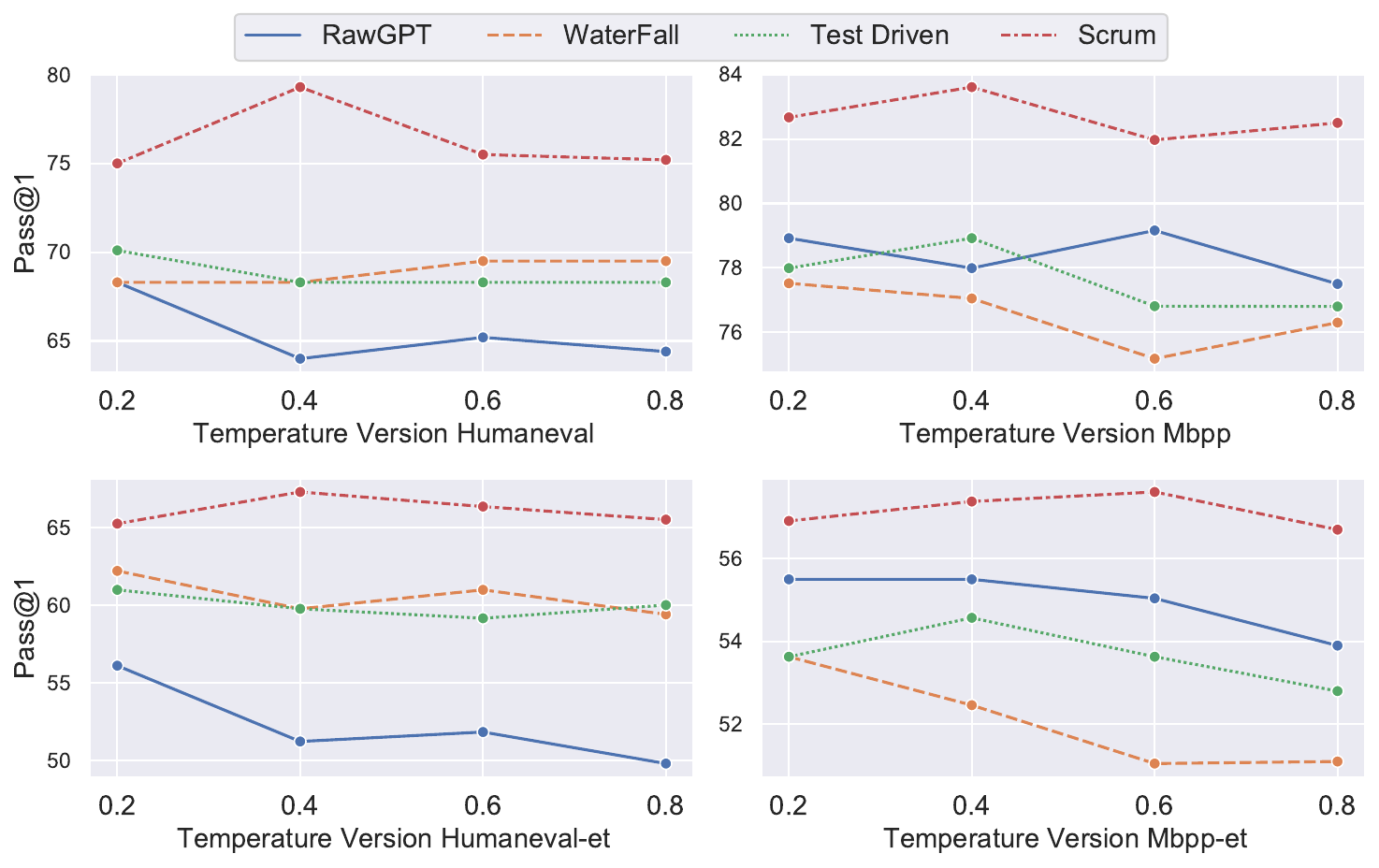}
  \caption{Pass@1 across temperature values. }
  \label{fig:rq2_temperature}
  \vspace{-6mm}
\end{figure}

\noindent{\underline{\textit{Result.}}} 
\noindent\textbf{\textit{\raw has extremely low Pass@1 in some versions of GPT3.5, while \tool has stable results across all versions. \tool may help ensure the stability of the generated code even when the underlying LLM regresses.}} Figure~\ref{fig:version} shows the Pass@1 for \raw and the three \tool across GPT3.5 versions. In earlier versions of GPT3.5 (\textit{0301} and \textit{0613}), \raw has very low Pass@1 on all benchmarks (e.g., 20 to 30 in HumanEval and HumanEval-ET). In \textit{0301}, MBPP's Pass@1 is even lower with a value around 5. \textbf{{The findings show that model version may have a significant impact on the generated code}}. 
However, we see that, after adopting our agent-based techniques, all three variants of \tool achieve similar Pass@1 across GPT3.5 versions. 
The results indicate that \tool can generate similar-quality code even if we have an underperformed baseline model. 

\noindent\textbf{\textit{All techniques have a relatively similar Pass@1 when the temperature value changes.}} Figure~\ref{fig:rq2_temperature} shows the Pass@1 for all the techniques when the temperature value changes. 
There is a slight downward trend for \raw when $t$ increases, but the changes are not significant (Pass@1 is decreased by 2 to 5). For \tool, and especially \toolScrum, we see similar Pass@1 regardless of the temperature value.
Although we see a slight increase in the Pass@1 of \toolScrum when $t$=0.4 (2 to 5 higher compared to when $t$=0.8 across the benchmarks), the difference is small, and the Pass@1 is almost the same when $t$ is either the lowest (0.2) or largest value (0.8). In short, although temperature values may have an impact on the generated code, the effect is relatively small\delete{, especially} for \tool.

\rqboxc{\tool generates stable results across GPT versions, while we see large fluctuations (14 times difference) in \raw's Pass@1. Pass@1 is generally consistent across all models when the temperature value changes. }
{
\subsection*{RQ4: How does \tool compare with other code generation techniques?}

\noindent{\underline{\textit{Motivation.}}}
\tool is designed to organize agents to emulate process models and can be combined with other code generation techniques. However, it is crucial to evaluate its performance relative to these techniques to assess the effectiveness in emulating software process models for code generation.
While many code generation results use the same benchmarks, our evaluation results cannot be directly compared with other agent-based code generation works~\cite{hong2023metagpt,huang2023codecot, huang2023agentcoder,chen2022codet,shinn2024reflexion} due to missing information on model versions, temperature values, post-processing steps, specific prompts, or the selection of few-shot samples. Therefore, in this RQ, we compare \toolScrum with other LLM-based baselines under the same environment settings. Moreover, we evaluate an integrated version of \toolScrum to showcase how existing prompting techniques can be combined with \tool. 

\noindent{\underline{\textit{Approach.}}} We compare against two state-of-the-art techniques: \codeT~\cite{chen2022codet} and \reflexion~\cite{shinn2024reflexion}. \codeT employs self-generated tests to evaluate the quality of generated code, which is similar to the testing phase of \tool. 
\reflexion is an agent-based technique that achieves state-of-the-arts Pass@1 on the benchmarks. 
We apply these two techniques using their released code, replacing the LLM version and temperature settings with those used by \toolScrum, and repeat the experiment five times. 

\noindent{\underline{\textit{Result.}}}
\textbf{\textit{While \toolScrum and \codeT achieve similar results, with \toolScrum having a higher Pass@1 in MBPP, they both have higher Pass@1 than \reflexion on all benchmarks (statistically significant).}}
Table~\ref{tab:rq4_accuracy} shows the Pass@1 of the techniques. \toolScrum has a statistically significantly higher Pass@1 than \codeT in MBPP and similar Pass@1 in other benchmarks (no statistically significant difference). Both techniques achieve higher Pass@1 than \reflexion (statistically significant). \reflexion's MBPP results are even worse than \raw. This observation aligns with the original study~\citep{shinn2024reflexion}, where Shinn et al. reported similar performance degradation. 

\noindent\textbf{\textit{Integrating \codeT to \toolScrum further brings statistically significant improvements of Pass@1 by up to 5\%.}} \codeT is a general technique where it repeats the code generation and selects the code that passes the most self-generated tests. Hence, as a pilot study, 
we made 
the developer agent repeats the implementation activity multiple times to produce several versions of the code (i.e., \toolScrumTest). The developer agent then generates multiple test assertions to identify the version with the highest pass rate. 
The selected code is subsequently submitted to the next stage: the testing activity. 
Our findings indicate that \toolScrumTest outperforms \toolScrum and \codeT, achieving an average Pass@1 score of 83.8, 58.3, 79.3, and 67.7 on HumanEval, HumanEval-ET, MBPP, and MBPP-ET, respectively. This provides 
statistically significant improvement over both \toolScrum and \codeT. 
Our finding highlights the potential of \tool in boosting the performance of other code generation techniques (and vice versa). Future studies can refine \tool to incorporate enhancement to each activity for further improvement. 
To support these efforts, we have made our code publicly available~\cite{dataset} to facilitate further adoption and allow researchers to experiment with different software process models. 

\rqboxc{Both \toolScrum and \codeT outperform \reflexion in Pass@1 across all benchmarks, with \toolScrum and \codeT demonstrating similar results. The incorporation of \codeT into \toolScrum further enhances performance, achieving the highest Pass@1 scores, highlighting the potential of \tool for code generation tasks.}

}

\begin{table}
	\caption{\add{Average and standard deviation of Pass@1 across five runs, with the best Pass@1 marked in bold.}}
	\label{tab:rq4_accuracy}
	\scriptsize
	\centering
 	\scalebox{0.85}{\setlength{\tabcolsep}{3.5mm}
	\begin{tabular}{lrrrr}
 
	\toprule	
 &HumanEval & HumanEval-ET & MBPP & MBPP-ET\\
	\midrule
        \reflexion & 71.3$\pm$1.5  & 55.7$\pm$2.8   & 71.7$\pm$0.8  & 52.0$\pm$0.7 \\
        \codeT & 75.7$\pm$0.4  & 66.9$\pm$0.4   & 79.9$\pm$1.3  & 56.7$\pm$0.9 \\
                
	\toolScrum & 75.2$\pm$1.1  & 65.5$\pm$1.9 & 82.5$\pm$0.6  &  56.7$\pm$1.4 \\
 \toolScrumTest & \textbf{79.3$\pm$1.6}  & {\bf 67.7$\pm$1.1} & {\bf 83.8$\pm$0.6} & {\bf 58.7$\pm$1.3} \\
	\bottomrule
	\end{tabular}
 	}
  \vspace{-3mm}

\end{table}

{
\section{Discussion \& Future Works}
\label{sec:discussion}

\phead{Role of Human Developers in \tool.} 
Although software process models were originally designed for human-centric development rather than for LLMs, our empirical findings suggest that certain elements of these processes can contribute to better code quality. Every activity in the process model also has different impacts on the generated artifacts. Future research should examine the incorporation of human developers into various phases of the code generation process. Specifically, humans can play critical roles in the following stages: \textbf{\textit{(1) Pre-Execution of \tool}}: different process models exhibit varying quality (e.g., smell and accuracy). Humans are instrumental in selecting the most appropriate model for a given task. Humans are also essential in providing initial requirements and design specifications. \textbf{\textit{(2) During \tool Execution}}: Humans can oversee review meetings and assist in reviewing/improving generated artifacts. For example, humans can validate the generated requirements with product managers, or verify the quality of the generated code by manual inspection and debugging. The improvement in each activity can also impact the subsequent activities and, hence, affect the final artifacts. \textbf{\textit{(3) Post-Execution of \tool}}: Following the code-generation phase, humans can either accept the generated artifact or request further refinements, offering additional requirements as needed to better meet project goals.

\phead{Quality of Code Generation Benchmarks}
We manually validated all coding problems that failed the tests. We found that the majority of quality issues within the benchmark were in MBPP and MBPP-ET (e.g., bad naming convention or inconsistent test definition). These issues may contribute to reduced Pass@1 scores due to factors beyond logic in the code. It is also important to acknowledge that other benchmarks might present unique challenges that could similarly affect Pass@1 evaluations. Hence, a crucial research direction is to conduct a thorough evaluation of benchmarks for more diverse and accurate evaluations of code generation approaches.}

\section{Threats to Validity}\label{threats}

\noindent\textbf{Internal validity.} Due to the generative nature of LLM, the responses may change across runs. \add{Variables such as temperature and LLM model version can also impact the generated code.} \add{We set the temperate value to be larger than 0 because we want LLMs to be more creative and offer more advice during discussions.} To mitigate the threat, we try to execute the LLMs multiple times. As found in RQ1, the standard deviation of the results is small, so the generated results should be consistent. 
\delete{Variables such as temperature and LLM model version can impact the generated code. To mitigate the threat, i}\add{I}n RQ3, we conducted the experiments using different temperature values and model versions. \delete{We find that t}\add{T}he temperature value only has a small effect on Pass@1, and model versions have a large impact on \raw. In RQ2, we study the impact of removing every development activity. However, having multiple development activities may have a tandem effect that further improves code quality. Future studies are needed to study the effects of different combinations of development activities in code generation. 

\noindent \textbf{External validity.} We conduct our study using two state-of-the-art benchmarks. However, as we discussed, there exist some issues in the provided tests. Moreover, the programming problems are mostly algorithmic, so the findings may not generalize to other code-generation tasks. Future studies should consider applying \tool on different and larger programming tasks. We use GPT3.5 as our underlying LLM. Although one can easily replace the LLM in our experiment, the findings may be different. Future studies on how the results of \tool change when using different LLMs. 

\noindent \textbf{Construct validity.} We try to implement an agent system that follows various software development processes. However, there are many variations of the same process model, and some variations may give better results. Future studies should further explore how changing the process models affect the code generation ability. 
\add{One limitation is that there is no guarantee of the correctness of the generated tests. However, we also found that the generated tests still contribute to improving the overall quality of the generated code. Similar findings are reported on CodeT~\cite{chen2022codet}, where generating multiple tests helps improve code correctness. However, future studies should focus on further improving the generated tests by using traditional software engineering techniques to estimate the oracles or select higher-quality tests (e.g., mutation testing).} 

\section{Conclusion}\label{conclusion}

\delete{Software development teams rely on following software process models to ensure software quality. Given that the Large Language Model (LLM) agents can act as domain experts and communicate with other agents, our study explores their efficacy in adopting engineering roles and facilitating interactions across various software process models.}
In this paper, we \add{emulate various roles in software development, such as requirement engineers, architects, developers, and testers, using Large Language Model (LLM) agents, and structuring their interactions according to established process models. We }introduce \tool, a framework that implements three renowned process models: \toolWater, \toolTdd, and \toolScrum. We evaluated how these models affect code generation in terms of correctness and code quality on four benchmarks: HumanEval, HumanEval-ET, MBPP, and MBPP-ET. Our findings show that \toolScrum notably enhances Pass@1 scores by an average of 15\% over \raw, while maintaining the lowest standard
deviation (averaging 1.2\%). 
\delete{While \raw's Pass@1 suffers significant variance when the model version changes (e.g., Pass@1 changes from 60 to 5 in HumanEval), the three process models still maintain stable Pass@1.} 
Moreover, \delete{we analyzed the generated code’s quality by examining code smells and the ability to handle exceptions. 
Our results suggest that incorporating }\add{we find that development activities such as }design and code review \delete{activities }significantly reduce code smells and increase the presence of handled exceptions. This indicates that \tool not only boosts code correctness but also reduces code smells and improves reliability. 
\add{Compared with other state-of-the-art techniques, \toolScrum and \codeT achieved similar
results, with both outperforming \reflexion. Integrating
\codeT into \toolScrum further resulted in statistically significant improvements, achieving the highest Pass@1 scores. }
These insights pave the way for future research to develop innovative development models tailored for LLM integration in software development processes.

In this study, we introduced \tool, a framework designed to emulate software process models using Large Language Model (LLM) agents, each representing roles such as requirement engineers, architects, developers, and testers. We implemented three variations of \tool: \toolWater, \toolTdd, and \toolScrum. Our evaluation across four benchmarks--HumanEval, HumanEval-ET, MBPP, and MBPP-ET--demonstrated the superior performance of \toolScrum, achieving up to 31.5\% improvement in Pass@1 scores over \raw.

Our results showed that incorporating software process models into LLM-based code generation significantly enhances code correctness, reduces code smells, and improves exception handling. FlowGenScrum consistently outperformed other models, achieving the highest Pass@1 scores and the lowest standard deviation, indicating more stable and reliable code generation.

Additionally, our comparative analysis with state-of-the-art techniques revealed that FlowGenScrum and CodeT achieved similar results, both outperforming Reflexion. Notably, integrating CodeT into FlowGenScrum resulted in statistically significant improvements, achieving the highest Pass@1 scores. This highlights the robustness and potential of combining structured software development practices with LLM capabilities.

Future research should focus on further refining these models, incorporating more sophisticated interactions, and expanding the scope of evaluation to include a broader range of software development tasks and environments. By doing so, we can better understand the capabilities and limitations of LLMs in software engineering and continue to improve their integration into practical development workflows.

These findings underscore the potential of LLMs in mimicking real-world development practices to generate higher-quality code and emphasize the importance of structured development practices in LLM-based code generation.

\balance

\bibliographystyle{plainnat}
\bibliography{Section/Main}

\end{document}